\documentclass[preprintnumbers,article,amsmath,amssymb,floatfix,14pt,prd,onecolumn,
superscriptaddress,nofootinbib]{revtex4-2}
\usepackage{bm}
\usepackage{amsfonts}
\usepackage{latexsym}
\usepackage[latin1]{inputenc}
\usepackage{graphicx}
\usepackage{amsmath}
\usepackage{palatino}
\usepackage{mathpazo}
\usepackage{textcomp}
\usepackage{float}
\usepackage{booktabs}
\usepackage{dcolumn}
\usepackage{ragged2e}
\usepackage{hyperref}
\hypersetup{colorlinks,citecolor=blue}
\hypersetup{colorlinks=true,linkcolor=red,filecolor=magenta,    urlcolor=blue}
\usepackage{amsmath}
\usepackage{xcolor}
\usepackage{orcidlink}
\usepackage{epsfig}
\usepackage{subfigure}
\usepackage{commath}

\def\jnl@style{\it}
\def\aaref@jnl#1{{\jnl@style#1}}

\def\aaref@jnl#1{{\jnl@style#1}}

\def\aj{\aaref@jnl{AJ}}                   % Astronomical Journal
\def\apj{\aaref@jnl{ApJ}}                 % Astrophysical Journal
\def\apjl{\aaref@jnl{ApJ}}                % Astrophysical Journal, Letters
\def\apjs{\aaref@jnl{ApJS}}               % Astrophysical Journal, Supplement
\def\apss{\aaref@jnl{Ap\&SS}}             % Astrophysics and Space Science
\def\aap{\aaref@jnl{A\&A}}                % Astronomy and Astrophysics
\def\aapr{\aaref@jnl{A\&A~Rev.}}          % Astronomy and Astrophysics Reviews
\def\aaps{\aaref@jnl{A\&AS}}              % Astronomy and Astrophysics, Supplement
\def\mnras{\aaref@jnl{Mon.~Not.~Roy.~Astron.~Soc.}}             % Monthly Notices of the RAS
\def\prd{\aaref@jnl{Phys.~Rev.~D}}        % Physical Review D
\def\prc{\aaref@jnl{Phys.~Rev.~C}}  % Physical Review C
\def\prl{\aaref@jnl{Phys.~Rev.~Lett.}}    % Physical Review Letters
\def\qjras{\aaref@jnl{QJRAS}}             % Quarterly Journal of the RAS
\def\skytel{\aaref@jnl{S\&T}}             % Sky and Telescope
\def\ssr{\aaref@jnl{Space~Sci.~Rev.}}     % Space Science Reviews
\def\zap{\aaref@jnl{ZAp}}                 % Zeitschrift fuer Astrophysik
\def\nat{\aaref@jnl{Nature}}              % Nature
\def\aplett{\aaref@jnl{Astrophys.~Lett.}} % Astrophysics Letters
\def\apspr{\aaref@jnl{Astrophys.~Space~Phys.~Res.}} % Astrophysics Space Physics Research
\def\physrep{\aaref@jnl{Phys.~Rep.}}      % Physics Reports
\def\physscr{\aaref@jnl{Phys.~Scr}}       % Physica Scripta
\def\commat{\aaref@jnl{Comm.~Math.~Phys.}}              % Communications in Mathematical Physics
\def\science{\aaref@jnl{Science}}               % Science
\def\cqg{\aaref@jnl{Classical Quant.~Grav.}}            % Classical and Quantum Gravity
\def\jpcs{\aaref@jnl{JPCS}}                                     % Journal of Physics Conference Series
\def\ijmpd{\aaref@jnl{Int.~J.~Mod.~Phys.~D}}                    % International Journal of Modern Physics D
\def\grg{\aaref@jnl{Gen.~Relat.~Gravit.}}               % General Relativity and Gravitation
\def\rpp{\aaref@jnl{Rep.~Prog.~Phys.}}          % Reports on Progress in Physics
\def\npa{\aaref@jnl{Nucl.~Phys.~A}}        % Nuclear Physics A
\def\lrr{\aaref@jnl{Living Rev.~Rel.}}                   % Living reviews in relativity
\def\jcap{\aaref@jnl{J.~Cosmology Astropart.~Phys.}}    % Journal of cosmology and astroparticle physics
\def\rmp{\aaref@jnl{Rev.~Mod.~Phys.}}   %Reviews of modern physics
\def\epjc{\aaref@jnl{Eur.~Phys.~J.~C}} 
\def\plb{\aaref@jnl{~Phy.~Lett.~B}} 
\def\mpla{\aaref@jnl{Mod.~Phy.~Lett.~A}} 
\def\arxiv{\aaref@jnl{arxiv.org}}

\allowdisplaybreaks[1]
\begin{document}
%\color{red}
\color{black}       %% For one column
\title{Dynamical system analysis of Dirac-Born-Infeld scalar field cosmology in coincident $f(Q)$ gravity}

\author{Sayantan Ghosh\orcidlink{0000-0002-3875-0849}}
\email{sayantanghosh.000@gmail.com}
\affiliation{Department of Mathematics, Birla Institute of Technology and Science-Pilani,\\ Hyderabad Campus, Hyderabad-500078, India.}

\author{Raja Solanki\orcidlink{0000-0001-8849-7688}}
\email{rajasolanki8268@gmail.com}
\affiliation{Department of Mathematics, Birla Institute of Technology and
Science-Pilani,\\ Hyderabad Campus, Hyderabad-500078, India.}

\author{P.K. Sahoo\orcidlink{0000-0003-2130-8832}}
\email{pksahoo@hyderabad.bits-pilani.ac.in}
\affiliation{Department of Mathematics, Birla Institute of Technology and Science-Pilani,\\ Hyderabad Campus, Hyderabad-500078, India.}
%\affiliation{Faculty of Mathematics \& Computer Science, Transilvania University of Brasov, Eroilor 29, Brasov, Romania}

%
%%%%%%%%%%%%%%%%%%%%%%%%%%%%%%%%%%%%%  DATE  %%%%%%%%%%%%%%%%%%%%%%%%%%%%%%%%%%%%
\date{\today}
\begin{abstract}
In this article, we offer the dynamical system analysis of the DBI (Dirac-Born-Infeld) scalar field in a modified $f(Q)$ gravity context. We have taken a polynomial form of modified gravity and used two different kinds of scalar potential, i.e., polynomial and exponential, and found a closed autonomous dynamical system of equations. We have analyzed the fixed points of such a system and commented on the conditions under which deceleration to late-time acceleration happens in this model. We have noted the similarity of the two models and have also shown that our result is indeed consistent with the previous work done on Einstein's gravity. We have also investigated the phenomenological implications of our models by plotting the EoS ($\omega$), Energy density ($\Omega$), and deceleration parameter ($q$) w.r.t. to e-fold time and comparing with the present value. Finally, we conclude the paper by observing how the dynamical system analysis differs in modified $f(Q)$ gravity, and we also provide some of the future scope of our work. 
\end{abstract} 

\maketitle

%\date{\today}

\textbf{Keywords:} DBI field, $f(Q)$ gravity, dark energy, dynamical system analysis.
%%%%%%%%%%%%%%%%%%%%%%%%%%%%%%%%%%%%%%%%%%%%%%%%%%%%%%%%%%%%%%%%%%%%%%%%
%%%%%%%%%%%%%%%        Introduction        %%%%%%%%%%%%%%%%%%%%%%%%%%%%%
%%%%%%%%%%%%%%%%%%%%%%%%%%%%%%%%%%%%%%%%%%%%%%%%%%%%%%%%%%%%%%%%%%%%%%%%
\section{Introduction}\label{sec1}
\justifying
After the discovery of CMB (Cosmic microwave background) in 1965 \cite{cmb}, it became clear that our universe started in a very hot, dense state called "Hot Big Bang" \cite{gamow} and has evolved in its current form. This is known as the standard Big Bang theory of cosmology. After the discovery of late time acceleration \cite{late1,late2} and observation from the galaxy rotation curve, it has been clear that there are other objects in our universe except for baryonic matter \cite{planck}. In the standard $\Lambda$CDM paradigm, which is probably the most successful theory about the current state of the universe, one takes dark energy (which is responsible for late time acceleration) as Cosmological constant $\Lambda$ and dark matter to be cold (non-radiative). Even though the $\Lambda$CDM model is so successful with phenomenological predictions and observational evidence, it has some severe problems. One of the main problems is the nature of dark energy. If one assumes that cosmological constant ($\Lambda$) is solely responsible for dark energy, then calculations from QFT can be shown to have a discrepancy of order ($10^{120}$) \cite{weinberg}. One natural way to explain this is by introducing the scalar field (Quintessence field \cite{carroll}), which can explain why the current value of the cosmological constant is so low. The scalar field also appears quite naturally in the early inflation scenarios, which can naturally explain the horizon problem and flatness problem, etc.  \\
Even though the scalar field can explain both early inflation and late-time acceleration, the exact form or the origin of the scalar field is not known. There are many such candidates for the origin of the inflation or quintessence fields. In this article, we take Dirac-Born-Infeld (DBI) as the origin of the scalar field, which naturally comes from the string theory. We have also done a dynamical system analysis in the flat FLRW background and given the phenomenological predictions (evolution graph of $q$, $\Omega$, $\omega$) based on the fixed point analysis.\\
It is well known that Einstein's general theory of relativity is not renormalizable in the context of quantum field theory. There have been several attempts to find a renormalizable theory of quantum gravity, and string theory offers one such unification. It is well known that even in bosonic string theory, the quantization of the Polyakov action (conformal transformation of Nambu-Goto action) gives tachyon like a field which soon decays via spontaneous symmetry breaking \cite{green,polchinski}. It was first observed by Mazumdar et al. \cite{mazumdar} that the decay of a non-BPS $D4$ branes to a stable $D3$ branes can give rise to tachyon field, which can act as an inflation field in the cosmological context. In 2002, a series of three papers by Sen \cite{sen1,sen2,sen3} showed how, in string theory, as well as in string field theory, tachyons occur naturally, and in \cite{sen3} it has been shown that the effective field of such tachyons can be viewed as DBI scalar field theory.\\
Soon after these proposals, Padmanabhan \cite{paddy} and Gibbons \cite{gibbons1} showed how these DBI-type fields could be used in the FLRW background to give inflation field-like behaviour. Alternative ways of getting the DBI field from other forms of string theory have been reviewed by Gibbons \cite{gibbons2}. The study of the DBI field in late time acceleration context has been done by Bhagla et al. \cite{bhagla}, while Gorini et al.\cite{gorini}
offered the alternative way of visualizing the DBI field as a modified Chaplygin gas. We also note that DBI field has been proposed as an alternative to dark matter by Padmanabhan \cite{paddy2} this shows that the DBI field could indeed effect the late time cosmology.\\
 Copeland \cite{copeland1} and Aguirregabiria \cite{aguirregabiria} first studied the study of the DBI field in the dynamical system setting. In this paper also, we are closely following the treatment given in \cite{copeland1}. Soon after that, Fang and Lu \cite{fang2010a} considered a much more general type of potential beyond inverse square potential, the work later extended by Quiros et al. \cite{Quiros} to include much more general potentials, and they have given an exact treatment of $sinh(\phi)$ potential. Guo \cite{guoexp} has chosen an exponential potential for dynamical system analysis which we have used here to make an autonomous dynamical system. It is also worth noting that as Silverstein and Tong \cite{tong} have shown, if one considers a D3-brane moving towards the horizon of AdS space, one can get a generalized DBI field in a strong coupling limit (as opposed to a weak coupling limit where the previous work has been done). In strong coupling limit, it can shown that DBI field gets extra contributions from the movement of the D3-brane and the lagrangian becomes $\mathcal{L}_{GDBI}=\frac{1}{f(\phi)}(\sqrt{1+f(\phi)\partial \phi^2}-1)-V(\phi)$.\\
The conventional concept of relativity, especially General Relativity, which interprets gravity as the curvature of spacetime, may not offer the definitive solution to elucidate dark energy. This encourages the exploration of alternative theoretical frameworks in cosmology that can effectively address cosmic acceleration while remaining consistent with observational data. General Relativity and its curvature-based extensions have been formulated and thoroughly examined in previous research \cite{R14, R15}. Recently, alternative theories of gravitation based on a flat spacetime geometry, relying solely on non-metricity, have been established and extensively explored \cite{R16, R17}. The $f(Q)$ gravity, with its various astrophysical and cosmological implications, has been widely investigated \cite{R18, R19, R20, R21, R22, R23, R24, R25, R26, R27, R28, R29}. In this article, we have shown that even with the modified $f(Q)$ gravity, we are getting a similar kind of late-time accelerating behavior where $q$ is $-1$ as expected from de-Sitter-like expansion. We also note that in our investigation, the present value of the deceleration parameter is coming to nearly $-0.8$, which is quite consistent with the observed value $-0.55$.\\
We initiate our exploration by introducing a set of dimensionless variables that encapsulate the complete evolution of the system's phase space. These variables facilitate the transformation of the system's dynamics into an autonomous structure, thereby enhancing our comprehension of the system's behavior. Several noteworthy findings within the context of modified gravity utilizing the dynamical system techniques have been appeared in references \cite{DE-3,WOM,HAMID,APLL,OO1,OO2}. We note that in this article, we have taken both DBI scalar field (for quintessence field )and $f(Q)$ gravity to discuss the late time acceleration phase. We are taking the DBI field as a quintessence field to explain the late time acceleration, which is reasonable as the inflationary field could indeed be responsible for late time acceleration as discussed in \cite{peeble1,peeble2}. Also, in those energy scales, it is reasonable to expect that $f(Q)$ gravity would emerge as an effective field theory of higher order correction to graviton graviton interactions \cite{effective}. The general criteria for the DBI field to give de-Sitter-like later time acceleration is given by the theorem of Hao\cite{hao} and Chingangbam \cite{chingangbam}. The present manuscript is organized as follows: In section \ref{sec2}, we present fundamentals of $f(Q)$ gravity formalism in the presence of scalar field, whereas motion equations corresponding to flat space-time have been presented in section \ref{sec3}. In section \ref{sec4}, we invoke the phase-space variables and perform the complete dynamical system analysis for the exponential and power-law potentials under the $f(Q)$ gravity formalism in the presence of DBI scalar field. Finally in section \ref{sec5}, we present our outcomes of the investigation.

\section{$f(Q)$ gravity in the presence of a scalar field}\label{sec2}
\justifying
It is well known that the original version of Einstein's equation using Riemannian geometry was written using the Levi-Civita connection. However, it was soon apparent that the connection in the Riemannian manifold could be more general than just Levi-Civita. It can be shown that a general connection can be broken into three different parts: Levi-Civita, anti-symmetric, and non-metricity. We refer the review article by Heisenberg \cite{Heisenberg/2024} for more details. In the most general form, the affine connection can be written in the form \cite{TRIN}:

\begin{equation}\label{2a}
\Upsilon^\alpha_{\ \mu\nu}=\Gamma^\alpha_{\ \mu\nu}+K^\alpha_{\ \mu\nu}+L^\alpha_{\ \mu\nu},
\end{equation}
Here the first term $\Gamma^\alpha_{\mu\nu}$ denotes the Levi-Civita connection,
\begin{equation}\label{2b}
\Gamma^\alpha_{\ \mu\nu}\equiv\frac{1}{2}g^{\alpha\lambda}(g_{\mu\lambda,\nu}+g_{\lambda\nu,\mu}-g_{\mu\nu,\lambda})
\end{equation}
The second term $K^\alpha_{\ \mu\nu}$ is a contortion tensor. The formula can written in the form of a torsion tensor ($T^\alpha_{\ \mu\nu}\equiv \Upsilon^\alpha_{\ \mu\nu}-\Upsilon^\alpha_{\ \nu\mu}$) as follows:
\begin{equation}\label{2c}
K^\alpha_{\ \mu\nu}\equiv\frac{1}{2}(T^{\alpha}_{\ \mu\nu}+T_{\mu \ \nu}^{\ \alpha}+T_{\nu \ \mu}^{\ \alpha})
\end{equation}
Finally, the last term is known as distortion tensor. The formula given in the form of the non-metricity tensor is given as follows:
\begin{equation}\label{2d}
L^\alpha_{\ \mu\nu}\equiv\frac{1}{2}(Q^{\alpha}_{\ \mu\nu}-Q_{\mu \ \nu}^{\ \alpha}-Q_{\nu \ \mu}^{\ \alpha})
\end{equation}	
The expression of the non-metricity tensor is given as,
\begin{equation}\label{2e}
Q_{\alpha\mu\nu}\equiv\nabla_\alpha g_{\mu\nu} = \partial_\alpha g_{\mu\nu}-\Upsilon^\beta_{\,\,\,\alpha \mu}g_{\beta \nu}-\Upsilon^\beta_{\,\,\,\alpha \nu}g_{\mu \beta}
\end{equation} 
We also define the superpotential tensor as follows:
\begin{equation}\label{2f}
4P^\lambda\:_{\mu\nu} = -Q^\lambda\:_{\mu\nu} + 2Q_{(\mu}\:^\lambda\:_{\nu)} + (Q^\lambda - \tilde{Q}^\lambda) g_{\mu\nu} - \delta^\lambda_{(\mu}Q_{\nu)}.
\end{equation}
where $Q_\alpha = Q_\alpha\:^\mu\:_\mu $ and $ \tilde{Q}_\alpha = Q^\mu\:_{\alpha\mu} $ are non-metricity vectors. If one contracts the non-metricity tensor with the superpotential tensor, one can get the non-metricity scalar ($Q$) as follows:
\begin{equation}\label{2g}
Q = -Q_{\lambda\mu\nu}P^{\lambda\mu\nu}. 
\end{equation}
The Riemann curvature tensor is given as follows:
\begin{equation}\label{2h}
R^\alpha_{\: \beta\mu\nu} = 2\partial_{[\mu} \Upsilon^\alpha_{\: \nu]\beta} + 2\Upsilon^\alpha_{\: [\mu \mid \lambda \mid}\Upsilon^\lambda_{\nu]\beta}
\end{equation} 
Now by using the affine connection \eqref{2a}, one can have
\begin{equation}\label{2i}
R^\alpha_{\: \beta\mu\nu} = \mathring{R}^\alpha_{\: \beta\mu\nu} + \mathring{\nabla}_\mu X^\alpha_{\: \nu \beta} - \mathring{\nabla}_\nu X^\alpha_{\: \mu \beta} + X^\alpha_{\: \mu\rho} X^\rho_{\: \nu\beta} - X^\alpha_{\: \nu \rho} X^\rho_{\: \mu\beta}
\end{equation}
Here, $\mathring{R}^\alpha_{\: \beta\mu\nu}$ and $\mathring{\nabla}$ are described in terms of the Levi-Civita connection \eqref{2b}. We also note that $X^\alpha_{\ \mu\nu}=K^\alpha_{\ \mu\nu}+L^\alpha_{\ \mu\nu}$. If one uses the contraction on Riemann curvature tensor using the torsion-free constraint  $ T^\alpha_{\ \mu\nu}=0$ in the equation \eqref{2i}, we get: 
\begin{equation}\label{2j}
R=\mathring{R}-Q + \mathring{\nabla}_\alpha \left(Q^\alpha-\tilde{Q}^\alpha \right)   
\end{equation}
Here, $\mathring{R}$ is the usual Ricci scalar evaluated regarding the Levi-Civita connection. We further use the teleparallel constraint, i.e., $R=0$. Upon using the teleparallel constraint and hence relation \eqref{2j} becomes:
\begin{equation}\label{2k}
\mathring{R}=Q - \mathring{\nabla}_\alpha \left(Q^\alpha-\tilde{Q}^\alpha \right)   
\end{equation}
From the above equation \eqref{2k}, we can see that the form of the Ricci scalar (using the Levi-Civita connection) differs from the non-metricity scalar ($Q$) by a total derivative. Using generalized Stoke's theorem, one can transform this total derivative into a boundary term. So we can see that the lagrangian density changes by a boundary term, and $Q$ is equivalent to $\mathring{R}$. As we can see, the $Q$ gives a comparable description of GR. We also note that as we have taken torsion to be zero, the theory is known as a symmetric teleparallel equivalent to GR (STEGR) \cite{KUHN}.\\
Now, we offer a general form of STEGR theory in the presence of a scalar field using a general form of $f(Q)$ in the lagrangian as follows:
\begin{equation}\label{2l}
\mathcal{S}=\int\frac{1}{2}\,f(Q)\sqrt{-g}\,d^4x+\int \mathcal{L}_{\phi}\,\sqrt{-g}\,d^4x\, ,
\end{equation}
where $g=\text{det}(g_{\mu\nu})$, $f(Q)$ is a function of the non-metricity scalar $Q$, and $\mathcal{L}_{\phi}$ denotes the Lagrangian density of a scalar field $\phi$ given by \cite{BAHA},

\begin{equation}\label{2m}
\mathcal{L}_{\phi} = -\frac{1}{2} g^{\mu \nu} \partial_\mu \phi  \partial_\nu \phi -V(\phi)
\end{equation}
Here, $V(\phi)$ is the potential for the scalar field. 
Now, by varying the above action \eqref{2l} with respect to the metric, we get the following field equation:
\begin{equation}\label{2n}
\frac{2}{\sqrt{-g}}\nabla_\lambda (\sqrt{-g}f_Q P^\lambda\:_{\mu\nu}) + \frac{1}{2}g_{\mu\nu}f+f_Q(P_{\mu\lambda\beta}Q_\nu\:^{\lambda\beta} - 2Q_{\lambda\beta\mu}P^{\lambda\beta}\:_\nu) = -T_{\mu\nu}^{\phi}.
\end{equation} 

Here, $f_Q=\frac{df}{dQ}$ and $T_{\mu\nu}^{\phi}$ is the energy-momentum tensor of the scalar field given as:
\begin{equation}\label{2o}
 T_{\mu\nu}^{\phi}= \partial_\mu \phi  \partial_\nu \phi -\frac{1}{2} g_{\mu \nu} g_{\alpha \beta} \partial^\alpha \phi  \partial^\beta \phi -  g_{\mu \nu} V(\phi)
\end{equation}
The scalar field satisfies the Klein-Gordon equation, which can be found by varying the action \eqref{2m} with respect to the $\phi$. The Klein-Gordon equation for the scalar field is given as follows:
\begin{equation}\label{2p}
\square \phi - V,_\phi =0
\end{equation}
Where $\square$ denotes the d'Alembertian and $V,_\phi = \frac{\partial V}{\partial \phi}$. \\
We also note by varying the action \eqref{2m} with respect to the connection (in the spirit of Palatini formulation) 
one can get the following equation:
\begin{equation}\label{2q}
\nabla_\mu \nabla_\nu (\sqrt{-g}f_Q P^{\mu\nu}\:_\lambda) =  0 
\end{equation}

\section{Equations of Motion}\label{sec3}
\justifying

In this article we assume that our universe is homogeneous and isotropic, which is evident from the large galaxy survey \cite{2df}. We would also like to note that the observation \cite{planck} suggests that the universe is also flat to a very good approximation, so the line element we are interested in is given by Friedmann-Lemaitre-Robertson-Walker (FLRW) metric. It can be shown \cite{hawking} that for homogeneous and isotropic Riemannian manifold, FLRW metric is the unique metric. Thus we consider the standard FLRW metric given by,
\begin{equation}\label{3a}
ds^2= -dt^2 + a^2(t)[dx^2+dy^2+dz^2]    
\end{equation}
Here, $a(t)$ is the scale factor of the universe's expansion. 
In the teleparallel consideration, we take the constraint corresponding to the flat geometry of a pure inertial connection. One uses a gauge transformation given by  $\Lambda^\alpha_\mu$  \cite{JIM-2}, to get the following form,
\begin{equation}\label{3b}
 \Upsilon^\alpha_{\: \mu \nu}  = (\Lambda^{-1})^\alpha_{\:\: \beta} \partial_{[ \mu}\Lambda^\beta_{\: \: \nu ]}
\end{equation}
We can also express the general affine connection by noting that a general element of $GL(4,\mathbb{R})$ characterized by the transformation $ \Lambda^\alpha_{\: \: \mu}=\partial_\mu \zeta^\alpha$, where $ \zeta^\alpha $ is an arbitrary vector field,
\begin{equation}\label{3c}
\Upsilon^\alpha_{\: \mu \nu} = \frac{\partial x^\alpha}{\partial \zeta^\rho} \partial_\mu \partial_\nu \zeta^\rho
\end{equation}
Due to gauge redundancy, one can eliminate the connection \eqref{3c} via a suitable coordinate transformation. Such a coordinate transformation is often called "gauge coincident". Using the coincident gauge we can calculate the on-metricity scalar corresponds to the metric \eqref{3a} becomes $Q=6H^2$.\\
The energy-momentum tensor for a perfect fluid distribution is given as:
\begin{equation}\label{3d}
T_{\mu\nu}=(\rho+p)u_\mu u_\nu + pg_{\mu\nu}
\end{equation}
where we have taken $u^{\mu}=(-1,0,0,0)$ are components of the four velocities. On comparing equation \eqref{3d} and \eqref{2o}, we have,
\begin{equation}\label{3e}
\rho=-\frac{1}{2}g_{\alpha \beta}\partial^\alpha \phi \partial^\beta \phi +V(\phi)
\end{equation}
\begin{equation}\label{3f}
p=-\frac{1}{2}g_{\alpha \beta}\partial^\alpha \phi \partial^\beta \phi -V(\phi)
\end{equation}
As GR does not depend on the coordinate choice, giving us the following expressions for pressure ($p_{\phi}$) and energy density ($\rho_{\phi}$) for the scalar field:
\begin{equation}\label{3g}
\rho_{\phi}=\frac{1}{2}\dot{\phi}^2+V(\phi)
\end{equation}
\begin{equation}\label{3h}
p_{\phi}=\frac{1}{2}\dot{\phi}^2-V(\phi)
\end{equation}
and the corresponding equation of state parameter can be written as,
\begin{equation}\label{3i}
\omega_{\phi}=\frac{p_{\phi}}{\rho_{\phi}}=\frac{\frac{1}{2}\dot{\phi}^2-V(\phi)}{\frac{1}{2}\dot{\phi}^2+V(\phi)}
\end{equation}
Also, in FLRW \eqref{3a} background, the Klein-Gordon equation \eqref{2p} takes the following form:
\begin{equation}\label{3j}
\ddot{\phi}+3H\dot{\phi}+V_{,\phi}=0
\end{equation}
From the field equation \eqref{2n} in FLRW background, in the presence of scalar field, we get the following Friendman-like equations:
\begin{equation}\label{3k}
3H^2=\frac{1}{2f_Q} \left( -\rho_{\phi}+\frac{f}{2}  \right)
\end{equation}
\begin{equation}\label{3l}
    \dot{H}+3H^2+ \frac{\dot{f_Q}}{f_Q}H = \frac{1}{2f_Q} \left( p_{\phi}+\frac{f}{2} \right)
\end{equation}
We take $f(Q)$ functional as $f(Q)=-Q+\Psi(Q)$ (we note that one can get the ordinary GR by putting $\Psi=0$), we can rewrite the Friedmann equations \eqref{3k}-\eqref{3l} as following:
\begin{equation}\label{3m}
    3H^2= \rho_\phi + \rho_{de}
\end{equation}
\begin{equation}\label{3n}
    \dot{H}=-\frac{1}{2} [\rho_\phi + p_\phi+\rho_{de}+p_{de}]
\end{equation}
where $\rho_{de}$ and $p_{de}$ represent the energy density and pressure of the dark energy component, which contributes via the geometry of the spacetime,
\begin{equation}\label{3o}
    \rho_{de}=-\frac{\Psi}{2}+ Q\Psi_Q
\end{equation}
\begin{equation}\label{3p}
    p_{de}=-\rho_{de}-2\dot{H} \left( \Psi_Q+2Q\Psi_{QQ} \right)
\end{equation}

\section{The Dynamical System Analysis}\label{sec4}
\justifying
One of the main troubles of using string theory in cosmology directly is the so-called no-go theorem \cite{hao,chingangbam}, for wrapped products by compactifying the extra dimensions. We note that from the equations below the dynamical system equations, we can see that it does not get closed for the generalized DBI field but is closed for ordinary DBI. We also have compactified the phase space (the $\lambda$ axis) by using equation \ref{5g}. Using compactification, we have drawn the 3D phase space given in Figure \ref{fg1}. \\
In string theory, it was predicted by Sen \cite{sen1,sen2,sen3} there are 
tachyon fields in both open and closed string theory. For more on open and closed string theory, one can follow the reference \cite{polchinski}.  Even though for closed string theory, the tachyon fields are projected out in open string, they remain even though one can use a spontaneous symmetry-breaking argument to get rid of tachyon modes, one can still fully explain the reason for its existence. In the bosonic string theory, if one uses Nambu- Goto action then it is almost impossible to quantize in order to get meaningful quantization rules one has to invoke the conformal invariant Polyakov action, using the conformal field theory techniques one can quantize such an action which leads to the undesirable tachyon modes, even though they violate casualty it can be shown that they are unstable. So Tachyon modes are typically given by  Dirac-Born-Infeld (DBI) Lagrangian, which have the following form,
\begin{equation}\label{4a}
\mathcal{L}_{Tachyon}=V(\phi)\sqrt{1+\partial \phi^2}
\end{equation}
where $\partial\phi^2=\partial^{\mu}\phi\partial_{\mu}\phi$ and $V(\phi)$ is a potential function for the scalar field and $\partial^{\mu}\phi\partial_{\mu}\phi$ denotes the kinetic term for tachyon fields.\\
From the Lagrangian, one can find the field equation for the Tachyon field from the Euler-Lagrangian equation as,
\begin{equation}\label{4b}
    \frac{\Ddot{\phi}}{1-\dot{\phi}^2}+3H\dot{\phi}+\frac{V_{,\phi}}{V}=0
\end{equation}
This is the modified Klein-Gordon equation for the DBI field.\\
The Friedmann equation \eqref{3m}-\eqref{3n} becomes,
\begin{equation}\label{4c}
    3H^2= \rho_{DBI} + \rho_{de}
\end{equation}
\begin{equation}\label{4d}
    \dot{H}=-\frac{1}{2} [\rho_{DBI} + p_{DBI}+\rho_{de}+p_{de}]
\end{equation}
We note that for such cases the energy density ($\rho_{DBI}$) and pressure ($p_{DBI}$) are given by,
\begin{equation}\label{4e}
    \rho_{DBI}=\frac{V}{\sqrt{1-\dot{\phi}^2}}
\end{equation}
\begin{equation}\label{4f}
    p_{DBI}=-V\sqrt{1-\dot{\phi}^2}
\end{equation}
so the equation of state ($\omega_{DBI}$) is given by
\begin{equation}\label{4g}
    \omega_{DBI}=\frac{p_{DBI}}{\rho_{DBI}}=\dot{\phi}^2-1
\end{equation}
To construct the autonomous dynamical system, we use the following. We can define the variables as $x=\dot{\phi}$ and $y=\frac{\sqrt{V}}{\sqrt{3}H}$, so we can have $x^2=\dot{\phi}^2$ and $y^2=\frac{V}{3H^2}$ and $s^2=\Omega_{de}=\frac{\rho_{de}}{3H^2}$. Now, the equation \eqref{4c} becomes, 
\begin{equation}\label{4h}
 s^2 = 1-\frac{y^2}{\sqrt{1-x^2}}    
\end{equation}
We note that in order to form the dynamical system , it is more convenient to take the "e-folding" timing defined as $N=\ln a$ so we get $\frac{d}{dt}=H\frac{d}{dN} $\\
As $\dot{x}=\ddot{\phi}$ we can write $x^{\prime}=\frac{\Ddot{\phi}}{H}$ (where $\prime$ denotes the derivative with respect to "e-folding" time and $\dot{}$ denotes the derivative with respect to ordinary time). Utilizing this expressions in the Klein-Gordon equation for the DBI field given in \eqref{4b}, we get,
\begin{equation}\label{4i}
 \Ddot{\phi}= (1-x^2)[\lambda V^{\frac{1}{2}}-3Hx]    
\end{equation}
Here we have defined the variable $\lambda$ as $\lambda = -\frac{V_{,\phi}}{V^{\frac{3}{2}}}$.
Now using the equation \eqref{4i} and the fact that $y=\frac{\sqrt{V}}{\sqrt{3}H}$ in the expression $x^{\prime}=\frac{\Ddot{\phi}}{H}$, we obtain
\begin{equation}\label{4j}
 x^{\prime}=(x^2-1)(3x-\sqrt{3}\lambda y)   
\end{equation}
Further, on differentiating variable $y$ w.r.t e-folding time $N$, we obtain  
\begin{equation}\label{4k}
 y^{\prime}= -\frac{1}{2}y[\sqrt{3}\lambda xy+2\frac{\dot{H}}{H^2}]    
\end{equation}
Now, utilizing the equations \eqref{3o}-\eqref{3p} and \eqref{4e}-\eqref{4f} in the equation \eqref{4d}, we have
\begin{equation}\label{4l}
   \frac{\dot{H}}{H^2}= \frac{3x^2y^2}{2\sqrt{1-x^2}[\Psi_Q+2Q\Psi_{QQ}-1]}  
\end{equation}    
Hence, the equation \eqref{4k} becomes,
\begin{equation}\label{4m}
  y^{\prime}= -\frac{1}{2}y[\sqrt{3}\lambda xy+ \frac{3x^2y^2}{\sqrt{1-x^2}[\Psi_Q+2Q\Psi_{QQ}-1]}]   
\end{equation}
Now, in order to get the closed form of variable $\lambda$, we define another quantity $\Gamma$ as $\Gamma=\frac{VV_{,\phi\phi}}{V_{,\phi}}$. On differentiating the variable $\lambda$ w.r.t e-folding time $N$ with the quantity $\Gamma$, we obtain
\begin{equation}\label{4n}
  \lambda^{\prime}= \sqrt{3} xy\lambda^2[\frac{3}{2}-\Gamma]   
\end{equation}
In our analysis, we consider the cosmological model $f(Q)=-Q+\Psi(Q)=-Q+\alpha Q^n$ that have a great significance. A power-law correction to the STEGR will give rise to branches of solution applicable either to the early universe or to late-time cosmic acceleration. The model characterized by value $n < 1$ can describes late-time cosmology, potentially influencing the emergence of dark energy, whereas the model characterized by value $n > 1$ can describes the early universe phenomenon \cite{R20}.  Moreover, one can observe that the case $\alpha=0$ i.e. $\Psi=0 \Rightarrow f(Q)=-Q$ recovers the GR.\\
Hence by using $\Psi(Q)=\alpha Q^n$, we have $(\Psi_Q+2Q\Psi_{QQ}-1)=(2n-1)\alpha n Q^{n-1}-1$.\\
Moreover, $s^2=(-\frac{\Psi}{2}+Q\Psi_Q)\frac{1}{3H^2}= (2n-1)\alpha Q^{n-1}$. Thus, we have $[\Psi_Q+2Q\Psi_{QQ}-1]=ns^2-1 $. Therefore, the expression \eqref{4l} and \eqref{4m} becomes,
\begin{equation}\label{4o}
  \frac{\dot{H}}{H^2}= \frac{3x^2y^2}{[(n-1)\sqrt{1-x^2}-ny^2]}   
\end{equation}
\begin{equation}\label{4p}
  y^{\prime}= -\frac{1}{2}y[\sqrt{3}\lambda xy+ \frac{3x^2y^2}{[(n-1)\sqrt{1-x^2}-ny^2]}]    
\end{equation}
   
\subsection{Exponential potential}
\begin{figure}[H]
\centering
\includegraphics[scale=0.52]{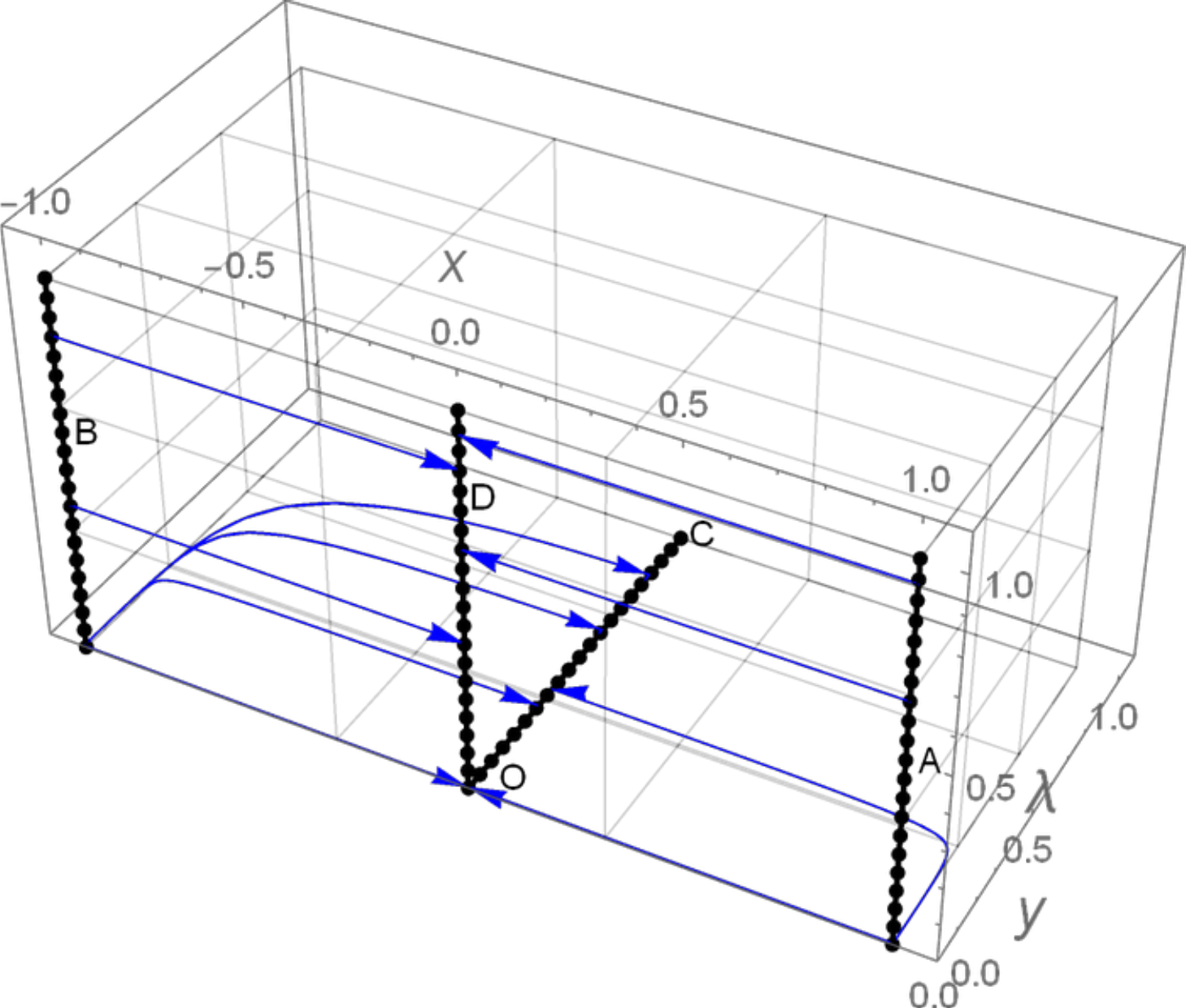}
\caption{The 3-D phase-space trajectories plotted for a set of solutions to the autonomous system \eqref{5b}-\eqref{5d} corresponding to the exponential potential.}\label{fg1}
\end{figure}

\begin{figure}[H]
{\includegraphics[scale=0.52]{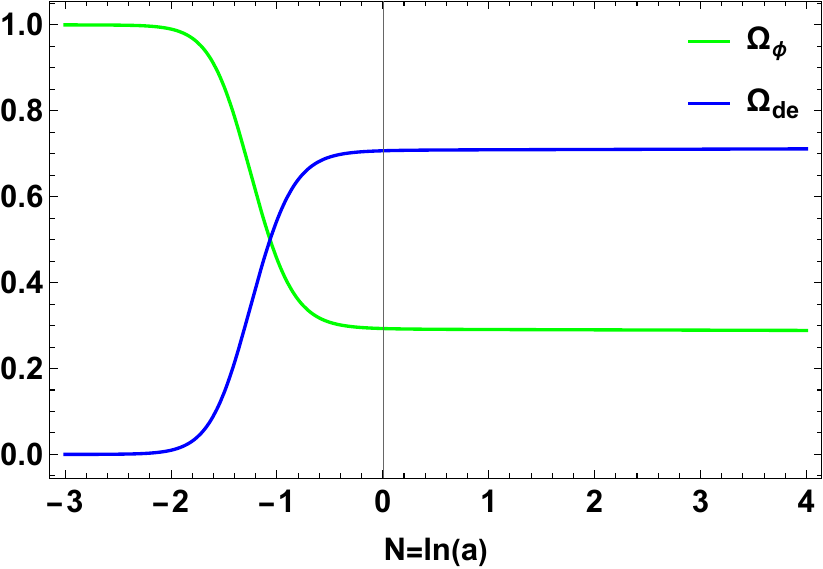}}
{\includegraphics[scale=0.56]{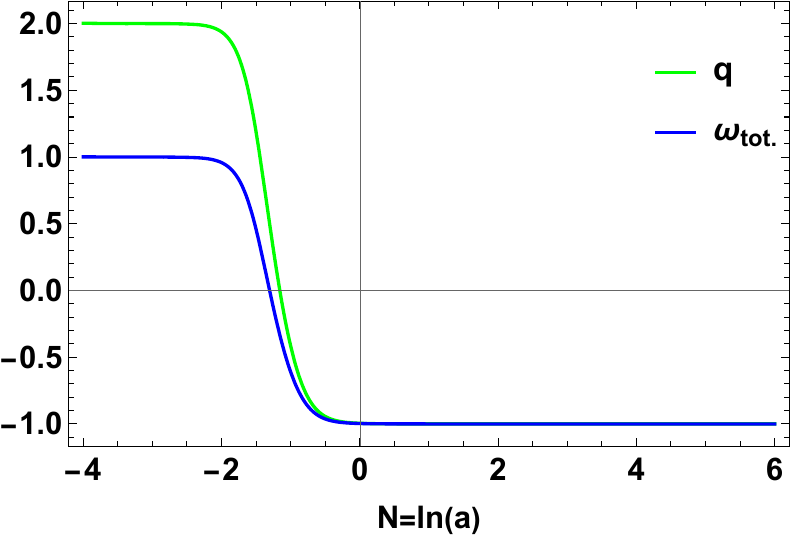}}
\caption{Evolutionary profile of scalar field density, dark energy density, deceleration, and the equation of state parameter for the exponential potential in DBI scalar field.}\label{fg2}
\end{figure}

\justifying
We note that exponential potential in DBI field can arise from the fact that if one takes the dark matter with the phantom field (which can naturally arise from the string theory) and applies the fact in the present epoch, the dark matter energy density and phantom energy density are comparable (coincidence problem). One can show that it would indeed give rise to exponential potential \cite{guoexp}. \\
Therefore, we take the following form of the exponential potential, 
\begin{equation}\label{5a}
V(\phi)= V_0 e^{-\beta \phi}
\end{equation}
Then for this choice of potential, we obtain $\lambda=-\frac{V_{,\phi}}{V^{\frac{3}{2}}}=\frac{\beta}{\sqrt{V_0e^{-\beta\phi}}}$, and $\Gamma=\frac{VV_{,\phi\phi}}{V_{,\phi}^2}=1$.\\
Therefore the complete autonomous form of dynamical equations \eqref{4j}, \eqref{4n}, and \eqref{4p} given as,
\begin{equation}\label{5b}
  x^{\prime}=(x^2-1)(3x-\sqrt{3}\lambda y)   
\end{equation}
\begin{equation}\label{5c}
  y^{\prime}= -\frac{1}{2}y[\sqrt{3}\lambda xy+ \frac{3x^2y^2}{[(n-1)\sqrt{1-x^2}-ny^2]}]   
\end{equation}
\begin{equation}\label{5d}
\lambda^{\prime}= \frac{\sqrt{3}}{2} xy\lambda^2  
\end{equation}
Utilizing the equation \eqref{4o} and the definition of deceleration parameter $q=-1-\frac{\dot{H}}{H^2}$, we obtain following expression corresponding to the model parameter $n=-1$,
\begin{equation}\label{5e}
q=-1-\frac{3x^2y^2}{-2\sqrt{1-x^2}+y^2}
\end{equation}
and the effective equation of state parameter is given by
\begin{equation}\label{5f}
\omega= \omega_{total}= \frac{p_{eff}}{\rho_{eff}}  = \frac{p_{DBI} + p_{de}}{\rho_{DBI} + \rho_{de}}= -1-\frac{2\dot{H}}{3H^2} = -1-\frac{2x^2y^2}{-2\sqrt{1-x^2}+y^2}
\end{equation}
We present the critical points and their behaviour (see Table \eqref{Table-1}) for the autonomous system \eqref{5b}-\eqref{5d} corresponding to the model parameter $n=-1$.
\begin{table}[H]
\begin{center}\caption{Table shows the critical points and their behaviour corresponding to the model parameter $n=-1$ and potential $V(\phi)= V_0 e^{-\beta\phi}$.}
\begin{tabular}{|c|c|c|c|c|c|}
\hline
 Critical Points $(x_c,y_c,z_c)$ & Eigenvalues ($\lambda_1$, $\lambda_2$,  $\lambda_3$) & Nature of critical point  & $q$ & $\omega$ \\
\hline 
$O(0,0,0)$ & $ (-3,0,0) $ & Nonhyperbolic (stable)  & $-1$ & $-1$ \\
$A(1,0,\lambda)$ & $ (6,0,0)  $ & Nonhyperbolic  & $-1$ & $-1$ \\
$B(-1,0,\lambda)$ & $ (6,0,0) $ & Nonhyperbolic  & $-1$ & $-1$ \\
$C(0,y,0)$ & $ (-3,0,0) $ & Nonhyperbolic (stable)  & $-1$ & $-1$ \\
$D(0,0,\lambda)$ & $ (-3,0,0) $ & Nonhyperbolic (stable)  & $-1$ & $-1$ \\
\hline
\end{tabular}\label{Table-1}
\end{center}
\end{table} 
We note that as $\lambda$ can take any value and is an even function (as the equation remains the same when $\lambda\rightarrow -\lambda$), we can take the physical region of the given dynamical system as the positive half cylinder with infinite length from $\lambda=0$ to $\lambda= +\infty$. Hence, we compactify the variable by defining the following phase-space variable $z$ as follows \cite{BAHA},
\begin{equation}\label{5g}
z=\frac{\lambda}{\lambda+1}  \:\: \text{or} \:\: \lambda=\frac{z}{1-z}
\end{equation}
Now, the evolutionary trajectories of the autonomous system \eqref{5b}-\eqref{5d} utilizing the above compactified variable is presented in the Figure \eqref{fg1}.
The evolutionary profile of scalar field density,
dark energy density, deceleration, and the equation of state parameter for the exponential potential have been presented in Figure \eqref{fg2}.
Note that, we have taken the entire plot in $ln(a)$ axis (see Figure \eqref{fg2}) and the present value of scale factor is taken to be $a=1$ that is $N=ln(1)=0$ is the present time. Further, the value $a < 1$ i.e. $N=ln(a) < 0$ represents distant past, whereas the value $a > 1$ i.e. $N=ln(a) > 0$ represents distant future.

\subsubsection{Limit point analysis for $(\pm1,0,\lambda)$}
Note that, for the critical point $A(1,0,\lambda)$ (obtained in Table \eqref{Table-1}), we find that the $q$ is taking an undefined form (see equation \eqref{5e}), so we are circumventing the problem by taking the appropriate limit of that fixed point.\\
 We first take $x=1-\epsilon_1$ and $y=\epsilon_2$ in the expression \eqref{4o} (for $n=-1$ case) where $\epsilon_1,\epsilon_2>0$ and when we will take $\epsilon_1,\epsilon_2\rightarrow 0$ we recover the original fixed points.
 \begin{equation}\label{5h}
\frac{\dot{H}}{H^2} =\frac{3(1-\epsilon_1)^2\epsilon_2^2}{\epsilon_2^2-2\sqrt{2\epsilon_1-\epsilon_1^2}}
     = \frac{3(1-\epsilon_1)^2}{1-2\sqrt{2\frac{\epsilon_1}{\epsilon_2^4}-\frac{\epsilon_1^2}{\epsilon_2^4}}}   
 \end{equation}
 If we happen to take the limit in such a way that $\frac{\epsilon_1}{\epsilon_2^4}\rightarrow 1$, that is $\frac{1-x}{y^4}\rightarrow1$, then as we note that $\frac{\epsilon_1^2}{\epsilon_2^4}\rightarrow0$ so get the following expression,
\begin{equation}\label{5i}
\frac{\dot{H}}{H^2}=\frac{3}{1-2\sqrt{2}}\approx -1.64<-1
\end{equation}
and hence we get,
\begin{equation}\label{5j}
q=-1-\frac{\dot{H}}{H^2}\approx .64
\end{equation}
We note that for a matter-dominated universe ($a=t^{\frac{2}{3}}$), we know $q=0.5$, as we can see in that our limit along that particular trajectory when $\frac{1-x}{y^4}\rightarrow1$ gives .64 which is quite consistent with the observation from matter dominated to late time acceleration phase.\\
In the same limit we also note that $\omega=-1-\frac{2x^2y^2}{-2\sqrt{1-x^2}+y^2}\approx0.09$\\
The deceleration parameter plays a vital role to describe the expansion phase of the universe. The value $q < 0$ depicts the accelerating behaviour whereas the transition from acceleration phase to deceleration phase admits the value $q \geq 0$. Thus,
\begin{equation}\label{5k}
q \geq 0 \iff  \frac{\dot{H}}{H^2} \leq -1 \iff  \frac{3}{2\sqrt{2\frac{\epsilon_1}{\epsilon_2^4}-\frac{\epsilon_1^2}{\epsilon_2^4}}-1} \geq 1 \iff 2 \geq \sqrt{2\frac{\epsilon_1}{\epsilon_2^4}-\frac{\epsilon_1^2}{\epsilon_2^4}} \iff  2 \geq\frac{\epsilon_1}{\epsilon_2^4} \iff 2 \geq \frac{1-x}{y^4}
\end{equation}
We note that in the above equations, we have used the fact $x\rightarrow1$ and $y\rightarrow0$. Also, the equality holds when $a\propto t$.\\
Similarly, the criteria \eqref{5k} gives $\omega\geq -\frac{1}{3}$. In addition, for arbitrary $n$, the criteria \eqref{5k} for the transition from acceleration phase to deceleration phase becomes $\frac{1-x}{y^4}\leq \frac{8}{(1-n)^2}$.
We also note that $\frac{\dot{H}}{H^2}=\frac{3}{1-2\sqrt{2}}$ gives $a\propto t^{\frac{2\sqrt{2}-1}{3}}\approx t^{0.609}$, we note that for matter dominated universe $a\propto t^{\frac{2}{3}}\approx t^{0.66}$. Similarly, for the case of $B(-1,0,\lambda)$ we can get a identical expressions with similar expression for $q$ and $\omega$, as the previous taking $\epsilon_1=x+1$ (noting that $x\geq-1$ and near $(-1,0,0)$).\\
\begin{figure}[H]
{\includegraphics[scale=0.39]{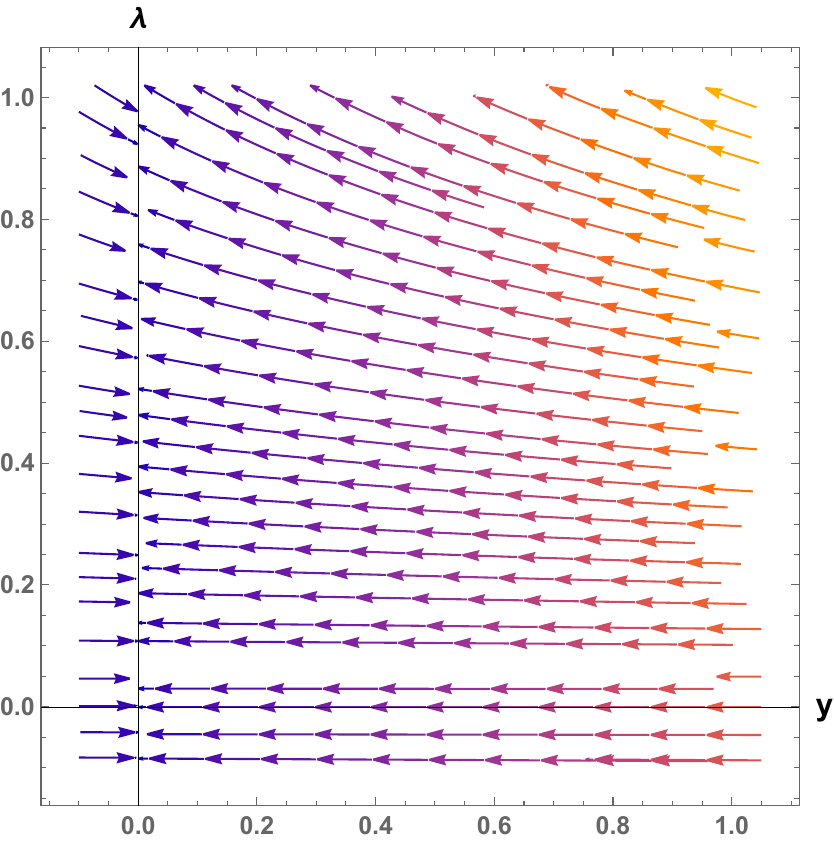}}
{\includegraphics[scale=0.39]{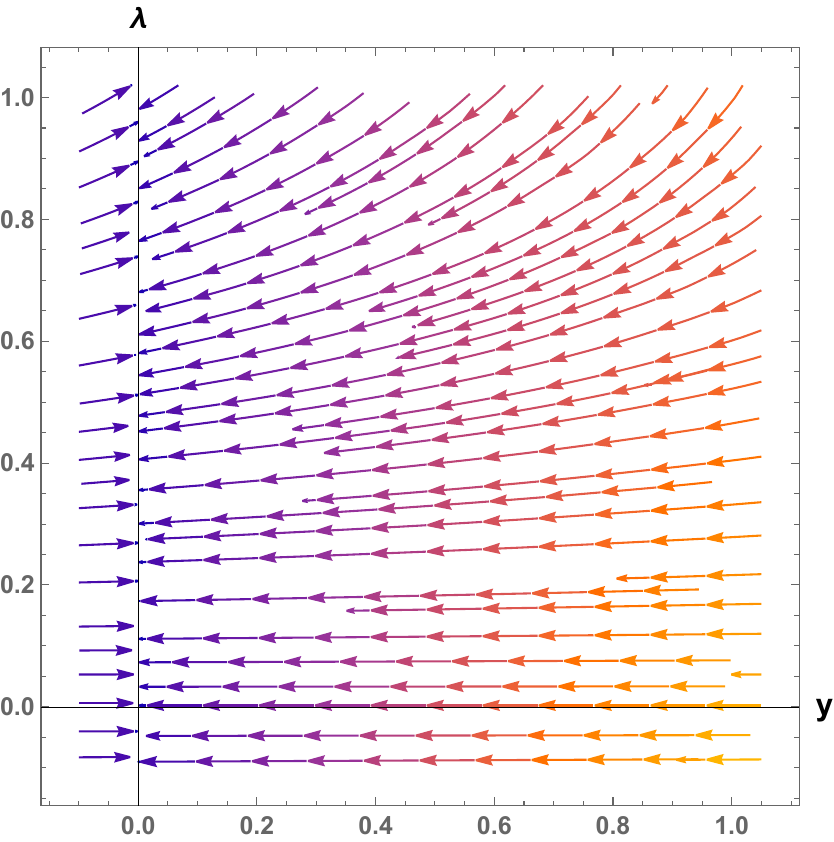}}
{\includegraphics[scale=0.55]{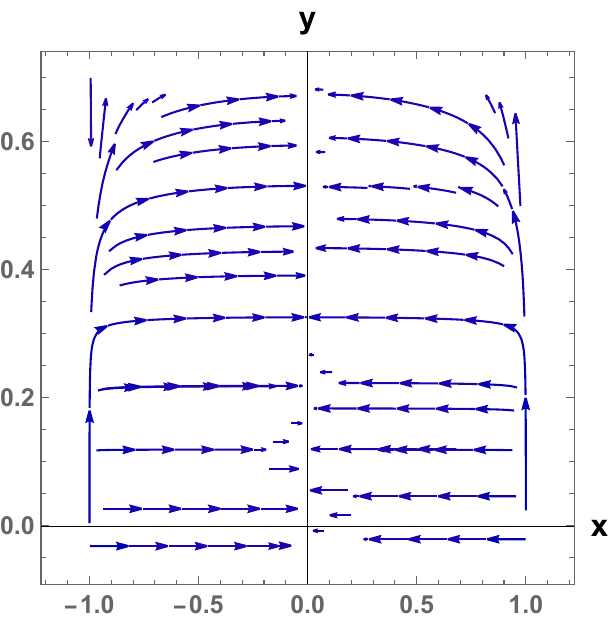}}
\caption{2D phase portrait for the $x=1,-1$ and $xy$ planes and shows that even though the critical points are non-hyperbolic, one can still take some particular limits to see the matter dominated to de Sitter transition.}\label{fg3}
\end{figure}
We see from Figure \eqref{fg3} that, we get a 2D phase portrait for $x=1$, as we can see that the corresponding eigenvalue for the critical point $A(1,0,\lambda)$, as we can see from the nature it is unstable, also $q=-1$ and $\omega=-1$ so it is giving a de Sitter type solutions. There is indeed a general theorem by \cite{hao} and \cite{chingangbam} that type of potential with a well-defined minimum would always lead to de-Sitter-type solutions. We can somewhat verify that by noting that the assertion holds true even in modified $f(Q)$ gravity.

\subsection{Power-law potential}
\begin{figure}[H]
\centering
\includegraphics[scale=0.52]{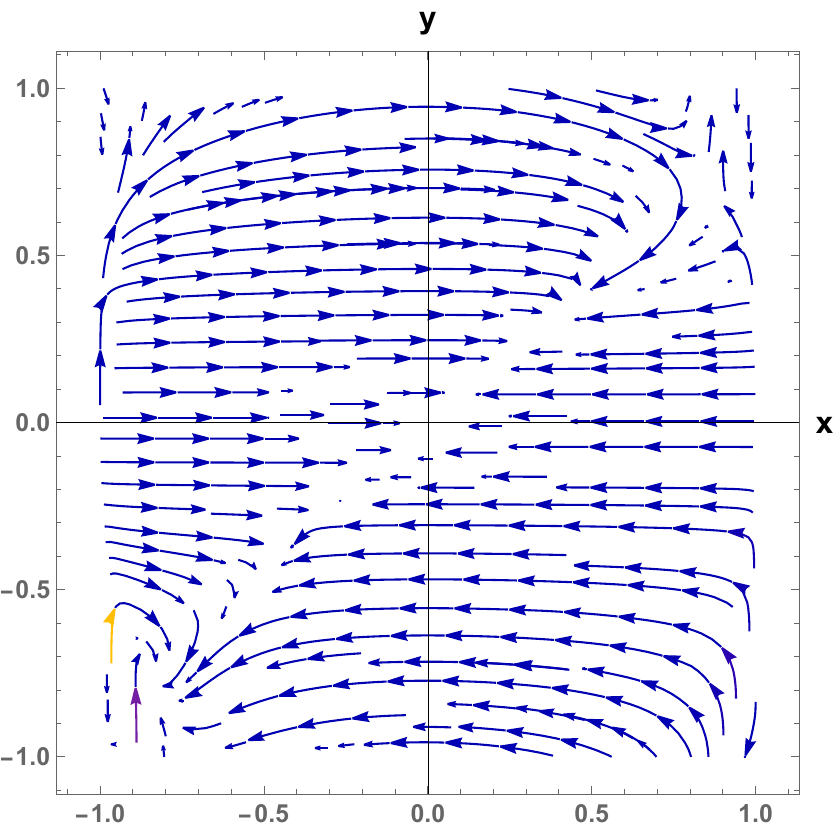}
\caption{2D phase portrait for polynomial potential when $k=2$}\label{fg4}
\end{figure}

\justifying
In this subsection, we take the power-law potential, noting that power-law potential most naturally gives global attractor solutions \cite{copeland1}. Also, it has been shown that these solutions are stable under perturbation \cite{fang2010a}.
We assume the following form of power-law potential,
\begin{equation}\label{6a}
V(\phi)= V_0\phi^{-k}
\end{equation}
Then for this choice of potential, we obtain, $\lambda=-\frac{V_{,\phi}}{V^{\frac{3}{2}}}=\frac{V_0k\phi^{-k-1}}{V_0\phi^{-k}(V_0\phi^{-k})^{\frac{1}{2}}}=\frac{k}{\sqrt{V_0}}\phi^{-1+\frac{k}{2}}$\\
In particular, for $k=2$ we get $\lambda=\frac{2}{\sqrt{V_0}}$.\\
Further, the quantity $\Gamma$ for the considered power-law potential becomes $\Gamma=\frac{VV_{,\phi\phi}}{V_{,\phi}^2}=\frac{k+1}{k}$,\\
Therefore the complete autonomous form of dynamical equations \eqref{4j}, \eqref{4n}, and \eqref{4p} for the power-law potential becomes,
\begin{equation}\label{6b}
x^{\prime}= (x^2-1)(3x-\sqrt{3}\lambda y)  
\end{equation}
\begin{equation}\label{6c}
y^{\prime}= -\frac{1}{2}y^2[\sqrt{3}\lambda xy+ \frac{3x^2y}{[(n-1)\sqrt{1-x^2}-ny^2]}]   
\end{equation}
\begin{equation}\label{6d}
\lambda^{\prime}= \frac{\sqrt{3}(k-2)}{2k} xy\lambda^2   
\end{equation}
We present the critical points and their behaviour (see Table \eqref{Table-2}) for the autonomous system \eqref{6b}-\eqref{6d} corresponding to the model parameter $n=-1$.
\begin{table}[H]
\begin{center}\caption{Table shows the critical points and their behaviour corresponding to the model parameter $n=-1$ and potential $V(\phi)= V_0\phi^{-k}$ with $k\neq2$.}
\begin{tabular}{|c|c|c|c|c|c|}
\hline
 Critical Points $(x_c,y_c,z_c)$ & Eigenvalues ($\lambda_1$, $\lambda_2$,  $\lambda_3$) & Nature of critical point  & $q$ & $\omega$ \\
\hline 
$O'(0,0,\lambda)$ & $ (-3,\sqrt{3}\lambda,0) $ & Stable (NH) for $\lambda<0$ and saddle for $\lambda\geq0$  & $-1$ & $-1$ \\
$A'(x,y,0)$ & $ (-3,0,0)  $ & Nonhyperbolic (stable)  & $-1$ & $-1$ \\
$B'(0,y,0)$ & $ (-3,0,0) $ & Nonhyperbolic (stable)  & $-1$ & $-1$ \\
\hline
\end{tabular}\label{Table-2}
\end{center}
\end{table}   
Note that the dynamical equations presented for the power-law case in \eqref{6b}-\eqref{6d} are identical to those presented for exponential case in \eqref{5b}-\eqref{5d}, they just differ by a constant in the $\lambda'$ equation, and hence the further analysis i.e. evolutionary trajectories are identical.\\
In particular, for the case $k=2$, the power-law case in \eqref{6b}-\eqref{6d} differs with the exponential one since it reduces to a following 2-dimensional dynamical system, as for the choice $k=2$ we have $\lambda'=0$.
\begin{equation}\label{6e}
x^{\prime}= (x^2-1)(3x-2\sqrt{3} y)  
\end{equation}
\begin{equation}\label{6f}
y^{\prime}= -\frac{1}{2}y^2[2\sqrt{3} xy+ \frac{3x^2y}{[(n-1)\sqrt{1-x^2}-ny^2]}]   
\end{equation}
It should be noted that it is not very surprising as it is shown in \cite{fang2010a} that $k=2$ indeed has a scale-invariant property which forces the it to become a two dimensional dynamical system of equations. 
Here, without loss of generality, we assume $V_0=1$ and hence $\lambda=\frac{2}{\sqrt{V_0}}=2$. We present the critical points and their behaviour (see Table \eqref{Table-3}) for the autonomous system \eqref{6e}-\eqref{6f} corresponding to the model parameter $n=-1$.
\begin{table}[H]
\begin{center}\caption{Table shows the critical points and their behaviour corresponding to the model parameter $n=-1$ with potential $V(\phi)= V_0\phi^{-k}$ having $k=2$ and $V_0=1$.}
\begin{tabular}{|c|c|c|c|c|}
\hline
 Critical Points $(x_c,y_c)$  & Nature of critical point  & $q$ & $\omega$ \\
\hline 
$O''(0,0)$ &  Stable  & $-1$ & $-1$ \\
$A''(0.806,0.698)$ & Saddle  & $0.362$ & $-0.092$ \\
\hline
\end{tabular}\label{Table-3}
\end{center}
\end{table}   
The evolutionary trajectories of the autonomous system \eqref{6e}-\eqref{6f} is presented in the Figure \eqref{fg4}.
From the phase-space trajectories obtained in the Figure \eqref{fg4}, it is evident that the solution trajectory indicates the evolution from the saddle point $A''$ representing matter-like behavior to the stable point $O''$ representing de-Sitter type accelerated expansion phase. Moreover, it which matches the previous analysis done by Copeland et al.\cite{copeland1}.

\section{Conclusion}\label{sec5}
\justifying

In this article, we have studied the DBI scalar field (the corresponding Lagrangian density has been presented in the equation \eqref{4a}) and its effect on cosmology via dynamical system analysis. The corresponding Klein Gordon equation with Friedmann like equations under the $f(Q)$ gravity formalism have been presented in equations \eqref{4b}, \eqref{4c}, and \eqref{4d}. For our analysis, we considered the cosmological model $f(Q)=-Q+\Psi(Q)=-Q+\alpha Q^n$ which is nothing but a power-law correction to the STEGR case, that will give rise to branches of solution applicable either to the early universe or to late-time cosmic acceleration. The model characterized by value $n < 1$ can describes late-time cosmology, potentially influencing the emergence of dark energy, whereas the model characterized by value $n > 1$ can describes the early universe phenomenon \cite{R20}. We obtained set of dynamical equations (i.e.\eqref{4j}, \eqref{4n}, and \eqref{4p}) corresponding to choice of our $f(Q)$ function. Further, in order to obtained the closed form (i.e. autonomous form) of the system, we investigated two specific forms of the potential function, specifically the exponential one $V(\phi)= V_0 e^{-\beta \phi}$ and the power-law $V(\phi)= V_0\phi^{-k}$, that have been extensively studied in the literature in GR context. We obtained the corresponding autonomous systems \eqref{5b}-\eqref{5d} and \eqref{6b}-\eqref{6d} and their stability analysis have been presented in Table \eqref{Table-1} and Table \eqref{Table-2}. Moreover, the 3-D phase-space trajectories of the autonomous system corresponding to exponential potential has been presented in the Figure \eqref{fg1} and the behavior of cosmological parameters such as deceleration, energy density, and effective equation of state parameter have been presented in the Figure \eqref{fg2}. Note that the autonomous system presented for the power-law case in \eqref{6b}-\eqref{6d} are identical to those presented for exponential case in \eqref{5b}-\eqref{5d}, they just differ by a constant in the $\lambda'$ equation, and hence the further analysis i.e. evolutionary trajectories are identical. We also observe that our fixed points have a late time de-Sitter type solution and that is somewhat we expected from the general theorem given by Hao et al. \cite{hao} and Chingangbam et al.\cite{chingangbam} about the sufficient condition for a de-Sitter type solution. Further, we analyzed the fixed point $(\pm1,0,\lambda)$ to show the transition from matter-dominated to de-Sitter type solutions. Also, we found the necessary and sufficient condition for the fixed points $(\pm1,0,\lambda)$ to show a matter-dominated phase to de-Sitter phase is given by $\frac{1-x}{y^4}\leq \frac{8}{(1-n)^2}$ (where in the limit $x\rightarrow 1$ and $y\rightarrow 0$) corresponding to the generic model parameter $n$. For the other fixed point $(-1,0,\lambda)$, one gets the similar results. In addition, we plotted the 2D phase portrait in Figure \eqref{fg3} for the $x=1,-1$ and $xy$ planes which shows that even though the critical points are non-hyperbolic, one can still take some particular limits to see the matter dominated to de-Sitter transition. We also note that in current time ($N=0$), we obtained $q\approx -0.8$ is somewhat consistent with the present observed value (-0.55). The discrepancy in $q$ value comes as there is no dark matter or ordinary matter in our calculations. We further note that the power-law potential case for $k=2$ is differ from the exponential one, since $k=2$ gives $\lambda^{\prime}=0$, making it a 2D autonomous system, that we have presented in \eqref{6e}-\eqref{6f}. The corresponding stability analysis and phase portrait are presented in the Table \eqref{Table-3} and Figure \eqref{fg4}. Moreover, the case $k=2$ matches the previous studies done in detail by \cite{bhagla,copeland1}. Thus, we conclude that the investigation presented in the manuscript successfully describes the late-time epochs of the universe, particularly de-Sitter expansion and the observed transition epoch along with effects of DBI field under the modified $f(Q)$ cosmology.\\

\section*{Data Availability}
There are no new data associated with this article.

\section*{Acknowledgments}
SG acknowledges the Council of Scientific and Industrial Research (CSIR), Government of India, New Delhi, for a junior research fellowship (File no.09/1026(13105)/2022-EMR-I). RS acknowledges the University Grants Commission (UGC), New Delhi, India, for awarding a Senior Research Fellowship (UGC-Ref. No.: 191620096030). PKS  acknowledges the Science and Engineering Research Board, Department of Science and Technology, Government of India, for financial support to carry out the Research project No.: CRG/2022/001847. We are very much grateful to the honorable referee and to the editor for the illuminating suggestions that have significantly improved our work in terms
of research quality, and presentation.

\end{document}